\documentstyle[12pt]{article}
\newcounter{mycount}

\begin{document}

\title{Exact Ground State Wave Functions for N Anyons in a Magnetic
Field on the Torus}
\author{Ansar Fayyazuddin \\ Institute of
Theoretical Physics \\ Stockholm University \\ Vanadisv\"{a}gen 9 \\
 S-113 46
Stockholm, Sweden}
\date{March 1993}
\maketitle

\abstract{ The complete set of ground state wave functions for N anyons
in an external magnetic field on the torus is found.  The cases
when the filling factor is less than or equal to one are considered.
The single valued description of anyons is employed through out by
coupling bosons to a Chern-Simons field.  At the end, the Chern-Simons
interaction is removed by a singular gauge transformation as a
result of which the wave
functions become multi-component in agreement with other
studies.}

\bigskip
USITP-93-07
\hfill
\newpage

Anyons \cite{lmw}, particles obeying fractional statistics, have been
the subject of intense study in the past few years.  Besides the purely
theoretical interest in fractional statistics a new motivation has
been provided by the Laughlin theory of the fractional quantum hall
effect (FQHE)\cite{lau}.  It was
realized some time ago that certain excitations of the  Laughlin state
are anyonic \cite{qph}.  Anyons are, therefore,
characteristic of the elusive Laughlin state.  An understanding
of anyons in a magnetic field may help to provide, in turn, an
understanding of the rich structure of the Laughlin state.  This
has been the motivation behind much of the study of anyons in magnetic
fields.  Some exact solutions on the plane can be found in ref \cite{mag}.

In this letter we study the problem of anyons in an external magnetic
field when the global topology is that of a
torus.  These, somewhat special, conditions put rather strict
constraints on the wave functions.
We will consider anyons on a torus which we parametrize by
a rectangle of dimensions $L_{1}\times L_{2}$ in the x-y plane with
opposite sides identified.  It is now known that anyons on the torus
can be equivalently studied by coupling bosons to a Chern-Simons
field.  The anyonic solutions can be recovered from the bosonic
theory by a singular gauge transformation.
The single valued description of anyons will be employed till the
end where we
recover the multi-valued, multi-component wave functions by performing
a singular gauge transformation.

The following notation will be used through out this letter:
$x_{\gamma}, y_{\gamma}$ denote the coordinates of particle
$\gamma$, and $X=\sum x_{\gamma}, Y=\sum y_{\gamma}$
the center of mass coordinates for the entire system.  The complex
coordinates  $z_{\gamma} \equiv \left(x_{\gamma} +
iy_{\gamma}\right)/L_{1}$ will sometimes be collected in a vector
${\vec z} \equiv z_{\gamma}{\hat e}_{\gamma}$, where the ${\hat
e}_{\gamma}$ denotes a vector with $N_A$ entries all of which are zero
except the $\gamma$th which is a one.  Use will also be made of the
vector ${\hat e} \equiv \sum_{\alpha = 1}^{N_A}{\hat e}_{\alpha}$. The
Chern-Simons coupling constant $\kappa = p/q > 0$, where $p,q$ are
relatively prime integers, is related to the statistical phase
$\theta$ in the anyon description by $\theta = \pi /\kappa$.   The
total flux $N_{\Phi} = \omega - N_{A}/\kappa$, where  $\omega =
BL_{1}L_{2}/2\pi$ and $N_A$ is the number of particles, is an integer
consistent with the Dirac quantization condition.  Notice that  we
need not require the Chern-Simons and magnetic flux to be integer
seperately since only the holonomy associated with the sum of the
gauge fields is an observable.  Finally, $N$ denotes the greatest
common denominator of $N_{\Phi}$ and $N_A$, $s\equiv N_{A}/N$,
$r\equiv N_{\Phi}/N$; needless to say, these definitions are only
sensible when $N_{\Phi}\neq 0$ which should be assumed unless
explicitly stated otherwise.  $a,b$ are defined by $s = ap+b$ with $0 \leq a$,
$0 \leq b < p$.  The modular parameter $\tau$ appearing in the theta
functions is defined to be $\tau = iL_{2}/L_{1}$.

The following solution to the Chern-Simons constraint will be used
\cite{ans} (other solutions are discussed in \cite{hh,il,salam}):
\begin{eqnarray}
a_{\alpha x} & = & \frac{\eta}{L_1} + \frac{2\pi}{\kappa
L_{1}L_{2}}Y - \frac{1}{\kappa}\sum_{\mu \neq \alpha}Im
\frac{\theta_{1} '\left(z_{\alpha}-z_{\mu}\mid \tau \right)}
{\theta_{1} \left(z_{\alpha}-z_{\mu}\mid \tau\right)}  \nonumber\\
a_{\alpha y} & = & \frac{\zeta}{L_2} - \frac{1}{\kappa}
\sum_{\mu \neq \alpha}Re  \frac{\theta_{1}
'\left(z_{\alpha}-z_{\mu}\mid \tau\right)} {\theta_{1}
\left(z_{\alpha}-z_{\mu}\mid \tau\right)}
\end{eqnarray}
$\eta$
and $\zeta$ are the part of the  Chern-Simons gauge field not
determined by the gauss law constraint generated by the Chern-Simons
term; they satisfy the commutation  relation:
\begin{equation}
\left[ \eta, \zeta \right] = i\frac{2\pi}{\kappa}
\end{equation}
Thus $\zeta$ can be represented as $\zeta =
-i\frac{2\pi}{\kappa}{\partial}/\partial_{\eta}$.  For the
external gauge field ${\vec A}$ the following gauge will be fixed:
\begin{equation}
{\vec A}_{\alpha} = -By_{\alpha} {\hat i}
\end{equation}

The Hamiltonian for this system is:
\begin{eqnarray}
H & = &-\frac{1}{2m}\sum_{\alpha = 1}^{N_A}{\vec D_{\alpha}}\cdot
{\vec D_{\alpha}} \nonumber \\
D_{\alpha x} & = &\frac{\partial}{\partial x_{\alpha}} - i\left[
\frac{\eta}{L_1} + \frac{2\pi }{\kappa L_{2}L_{1}}Y
-By_{\alpha} - \frac{1}{\kappa}\sum_{\mu \neq \alpha}Im
\frac{\theta_{1} '\left(z_{\alpha}-z_{\mu}\mid \tau \right)}
{\theta_{1} \left(z_{\alpha}-z_{\mu}\mid \tau\right)} \right]
\nonumber \\
D_{\alpha y} & = &\frac{\partial}{\partial y_{\alpha}} -
\frac{2\pi}{\kappa L_2}\frac{\partial}{\partial\eta} +
\frac{i}{\kappa}\sum_{\mu \neq \alpha}Re  \frac{\theta_{1}
'\left(z_{\alpha}-z_{\mu}\mid \tau\right)} {\theta_{1}
\left(z_{\alpha}-z_{\mu}\mid \tau\right)}
\end{eqnarray}
Since neither the Chern-Simons gauge field nor the external gauge
field are globally defined on the torus the hamiltonian is not
periodic under both cycles of the torus.  It is
easily checked that the gauge fields are periodic under $x_{\alpha}
\rightarrow x_{\alpha} + L_1$ but not under $y_{\alpha} \rightarrow
y_{\alpha} + L_2$.  Under the latter the hamiltonian changes by a gauge
transformation which is well defined on the torus if the total flux
$N_\Phi$
is an integer.  Whenever ${N_\Phi}\neq 0$, gauge invariant observables
are well defined only if the wave function changes as follows
\footnote{Only the coordinate indicated is transformed the others
are held fixed.}:
\begin{eqnarray}
\psi\left( x_{\alpha}+L_{1}\right) & = & \psi\left(x_{\alpha}\right)\\
\psi\left( y_{\alpha}+L_{2}\right) & = & e^{-i\frac{2\pi x_{\alpha}
N_{\Phi}}{L_1}} \psi\left(y_{\alpha}\right)
\end{eqnarray}
upto a possible global $U\left( 1 \right)$ rotation.  In addition
to translation invariance of observables, one has to impose
invariance under large gauge transformations \cite {il}: i.e.
covariance of the wave functions under $U_{1} = \exp
2\pi\partial_{\eta}  \exp-i2\pi X/L_{1}$ and $U_{2}= \exp
2\pi\partial_{\zeta}  \exp-i2\pi Y/L_{2} = \exp -i\kappa\left( \eta +
2\pi Y/\kappa L_2 \right)$.  The $U_i$ satisfy the following
commutation relation:
\begin{equation}
U_{1}U_{2} = U_{2}U_{1} e^{i2\pi\kappa}
\end{equation}
{}From now on we will work in a basis in which $U_2$ is diagonal, the
commutation relation then determines the action of $U_1$ upto a global
phase:
\begin{eqnarray}
U_{2}\psi_{l} & = &e^{-i\beta -i2\pi\kappa l} \psi_{l} \\
U_{1}\psi_{l} & = &e^{i\alpha} \psi_{l-1}
\end {eqnarray}
It should be emphasized that different components of the $q$ dimensional
representation of the group of large gauge transformations represent the same
physical state.  They are gauge copies of one another.

Having stated the quasi-periodicity
conditions for the wave functions in both space and the topological
part of the gauge field, we now seek eigenstates of the Hamiltonian
which satisfy these conditions.  First we expand the wave function in
the coordinates: $x_{\gamma}, y_{\gamma}$ and $v \equiv \eta + 2\pi
Y/\kappa L_2$.  Notice that the dependence on $\eta$ is
through $v$ only. Write $\psi_l$ as:
\begin{equation}
\psi_{l} = e^{-\frac{\pi\omega}{L_{1}L{2}}\sum_{\gamma}y_{\gamma}^{2}}
e^{vY/L_{1}}
\prod_{\gamma < \delta} \mid \theta_{1}\left( z_{\gamma} -
z_{\delta} \mid \tau \right) \mid^{\frac{1}{\kappa}} \phi_{l}
\left( x,y,v\right)
\end{equation}

Covariance under large gauge transformations and
$\left[ v, H \right] = 0$
allow for the following general expression for $\phi_l$:
\begin{equation}
\phi_{l}\left( x,y,v \right) = \sum_{n=0}^{p-1}\sum_{j=
-\infty}^{\infty} e^{i{\vec z}\cdot {\hat e}\left(l + jq\right)}
e^{i\alpha\left(l+jq + n/\kappa \right)} F_{n}\left( x,y\right)
\delta\left(\kappa v - 2\pi\kappa l -
\beta - 2\pi\left(jp+n\right)\right)
\end{equation}
Where the $F_{n}$ are bosonic functions independent of $v$.  Note that for
any holomorphic $F_n$ the $\psi_{l}$ are eigenstates of the
Hamiltonian with the lowest possible energy eigenvalue.

We define the magnetic translation operators:
\begin{eqnarray}
t_{\alpha}\left(m{\vec L_1}\right) & = &
e^{mL_{1}\frac{\partial}{\partial x}_{\alpha}}  \nonumber \\
t_{\alpha}\left( n{\vec L_2}\right) & = &e^{i2\pi nN_{\Phi}x_{\alpha}
/L_{1}}e^{nL_{2}\frac{\partial}{\partial y}_{\alpha}}
\end{eqnarray}
These operators commute among themselves and with the
Hamiltonian only when $m,n$ are integers.  When $N_{\Phi}$ vanishes
the translation operators commute among themselves for arbitrary
argument but not with the hamiltonian except in the singular limit in
which $p \rightarrow \infty$.
Quasi-periodicity under lattice translations is conveniently expressed
as invariance of the wave functions under these translation
operators for $m=n=1$.  The center of mass translation operators
\begin{eqnarray}
T\left({\vec L_i}\right) \equiv \prod_{\alpha}t_{\alpha}\left(
{\vec L_i}/N \right)
\end{eqnarray}
commute with the {$t_{\alpha}\left({\vec L_i}\right)$}, the
hamiltonian and among themselves.  This gives a maximal set of
translation operators that can be simultaneously diagonalized \cite{hal}.
Since we are dealing here with a system of identical particles
$t_{\alpha}=t_{\beta}$ on the Hilbert space of wave functions.
Explicitly, magnetic translation invariance
\begin{eqnarray}
t_{\alpha}\left({\vec L}_{i}
\right)\psi_{l}\left(\left\{x,y\right\},v\right) =
\psi_{l}\left(\left\{x,y\right\},v\right)
\end{eqnarray}
imposes the following conditions on the holomorphic
$F_{n}\left({\vec z}\right)$:
\begin{equation}
F_{n}\left({\vec z} + {\hat e}_{\gamma}\right)  = F_{n}\left({\vec
z}\right)
\end{equation}
\begin{equation}
e^{i2\pi N_{\Phi}{\hat e}_{\gamma} \cdot{\vec z} +
i\pi\tau\left(N_{\Phi}+\frac{1}{\kappa}\right)
-i\frac{\tau}{\kappa}\left(2\pi n + \beta\right) +
i\frac{\alpha}{\kappa}}  F_{n}\left({\vec z} + \tau{\hat
e}_{\gamma}\right) = F_{n-1}\left({\vec z}\right) \mbox{ (for $n>0$) }
\end{equation}
\begin{equation}
e^{i2\pi \left(N_{\Phi}{\hat e}_{\gamma}+q{\hat e}\right)
\cdot{\vec z} +
i\pi\tau\left(N_{\Phi}+\frac{1}{\kappa}\right)
-i\frac{\tau}{\kappa}\beta +
i\frac{\alpha}{\kappa}}  F_{0}\left({\vec z} + \tau{\hat
e}_{\gamma}\right) = F_{p-1}\left({\vec z}\right)
\end{equation}
By repeated application of $t_{\alpha}\left({\vec L}_{2}
\right)$ the following relation can be derived:
\begin{eqnarray}
e^{i2\pi \left(pN_{\Phi}{\hat e}_{\gamma}+q{\hat e}\right)
\cdot{\vec z} +
i\pi\tau p^{2}\left(N_{\Phi}+\frac{1}{\kappa}\right)
-ip \frac{\tau}{\kappa}\left(2\pi n +\beta\right) +
iq\alpha}  F_{n}\left({\vec z} + p\tau{\hat
e}_{\gamma}\right) & = & F_{n}\left({\vec z}\right)
\end{eqnarray}

The next step in constructing a complete set of functions $F_n$ is
best motivated by pausing here to state some definitions and a
theorem \cite{mum} of which use will be made later.

Definition: Given $\Omega$, a $g\times g$ complex, symmetric matrix
such that $Im \Omega$ is positive definite (in the mathematical
literature the space of such matrices ${\cal M}_g$ is called
the Siegel upper-half-space), we define the theta function
$\theta\left[\begin{array}{c}{\vec a}\\{\vec
b}\end{array}\right]\left({\vec z},\Omega\right)$ with characteristics
$\left({\vec a},{\vec b}\right)$ as:

\begin{equation}
\theta\left[\begin{array}{c}{\vec a}\\{\vec
b}\end{array}\right]\left({\vec z},\Omega\right)
= \sum_{{\vec n} \in Z^g}
\exp \left[i\pi\left({\vec n} + {\vec a}\right)\cdot
\Omega\left({\vec n} + {\vec a}\right) + i2\pi
\left({\vec n} + {\vec a}\right)\cdot\left({\vec z}+{\vec b}\right)
\right]
\end{equation}
The positive definiteness of $Im \Omega$ guarantees uniform
and absolute convergence of the series defining the theta function.
These functions satisfy the following quasi-periodicity conditions:
\begin{equation}
\theta\left[\begin{array}{c}{\vec a}\\
{\vec b}\end{array}\right]\left({\vec z} +{\vec m},
\Omega\right)  =  e^{i2\pi {\vec a}\cdot{\vec m}}
\theta\left[\begin{array}{c}{\vec a}\\{\vec
b}\end{array}\right]\left({\vec z},\Omega\right)
\end{equation}
\begin{equation}
\theta\left[\begin{array}{c}{\vec a}\\{\vec
b}\end{array}\right]\left({\vec z} +\Omega{\vec m},
\Omega\right) = e^{-i2\pi {\vec b}\cdot{\vec m}}
\exp \left(-i\pi{\vec m}\cdot
\Omega{\vec m} - i2\pi
{\vec m}\cdot{\vec z}\right)
\theta\left[\begin{array}{c}{\vec a}\\{\vec
b}\end{array}\right]\left({\vec z},\Omega\right)
\end{equation}
\begin{equation}
\theta\left[\begin{array}{c}{\vec a}+{\vec m}\\{\vec b}+{\vec n}
\end{array}\right] \left({\vec z}, \Omega\right) = e^{i2\pi {\vec
a}\cdot{\vec n}} \theta\left[\begin{array}{c}{\vec a}\\{\vec
b}\end{array}\right]\left({\vec z},\Omega\right)
\end{equation}
whenever ${\vec m},{\vec n} \in Z^{g}$.

Definition: Fix $\Omega \in {\cal M}_g$.  Then an entire function
$f\left({\vec z}\right)$ on $C^g$ is $L_{\Omega}$-quasi-periodic
of weight $\ell$ if
\begin{eqnarray}
f\left({\vec z}+{\vec m}\right) & = & f\left({\vec z}\right) \\
f\left({\vec z}+\Omega{\vec m}\right) & = &
\exp \left(-i\pi \ell{\vec m}\cdot{\Omega}{\vec m}
-i2\pi \ell{\vec m}\cdot{\vec z}\right)
f\left({\vec z}\right)
\end{eqnarray}
for all ${\vec m} \in Z^g$.  Let $R^{\Omega}_{\ell}$ be the vector
space  of such functions $f$.

Theorem: Fix $\Omega \in {\cal M}_g$.  Then a basis of
$R^{\Omega}_{\ell}$ is given by:
\begin{equation}
f_{{\vec c}} = \theta\left[\begin{array}{c}{\vec 0}\\ \frac{{\vec
c}}{\ell} \end{array}\right]\left({\vec z},\ell^{-1}\Omega\right)
\end{equation}
for $c_{i} \in Z$, $0 \leq c_{i} < \ell$.

Though this theorem is very powerful its proof is simple \cite{mum}: it
immediately follows by noting that the first condition allows
one to expand the function in a Fourier series and that the second
condition gives a recursion relation between the coefficients of the
series. The most general solution for the coefficients establishes the
theorem. Of course a different representation for the basis elements is
always possible.

Returning to the problem at hand, we would like to find a basis of
functions $F_n$ by using the above theorem.  To this end we rewrite
the quasi-periodicity relation for the $F_n$ as:
\begin{eqnarray}
& &F_{n}\left({\vec z}+p\tau{\hat e}_{\gamma}\right) =
F_{n}\left({\vec z}+\Omega '{\vec m}_{\gamma}\right) \nonumber \\
& = &\exp \left(-i\pi \ell{\vec m}_{\gamma}\cdot{\Omega '}{\vec
m}_{\gamma} -i2\pi \ell {\vec m}_{\gamma}\cdot\left({\vec
z}-\frac{\tau}{\omega\kappa}\left(n+\frac{\beta}{2\pi}\right){\hat e}
+\frac{\alpha}{2\pi\omega\kappa}\right)\right)F_{n}\left({\vec
z}\right)
\end{eqnarray}
Comparing with equation (18) gives:
\begin{equation}
\ell{\vec m}_{\gamma} = pN_{\Phi}{\hat e}_{\gamma} + q{\hat e}
\end{equation}
and
\begin{equation}
\Omega '= \frac{\ell\tau}{\omega\kappa N_{\Phi}}
\left( \begin{array}{c} \left(\omega\kappa-1\right) \mbox{  }-1 \cdots
-1 \\ \\ \ddots \\
-1 \cdots -1 \mbox{  }\left(\omega\kappa-1 \right)\end{array} \right)
\end{equation}
$Im \Omega '$ is positive definite only if $\omega\kappa - N_{A} > 0$
which we assume from now on.  This is essentially the condition
that the anyonic generalization of the Landau level filling
factor, $N_{A}/ \omega\kappa$, be less than one (the case where
$\omega\kappa = N_A$ will be dealt with seperately below).
To fix $\Omega '$, require that for any ${\vec n}\in Z^{N_A}$
\begin{equation}
\tau^{-1}\Omega '{\vec n} \in pZ^{N_A}
\end{equation}
The choice $\ell= p^{2}\omega N_{\Phi}$ fullfills the above condition
and $F_n$ belongs to $R^{\Omega '}_{\ell}$.  The
theorem then provides us with a basis of $R^{\Omega '}_{\ell}$.  Using
these functions we can construct the $F_n$ which satisfy rather more
stringent conditions than a general member of $R^{\Omega '}_{\ell}$.  A
basis for $R^{\Omega '}_{\ell}$ is given by the
set of functions:
\begin{equation}
g_{n,{\vec c}} = \theta\left[\begin{array}{c}{\vec 0} \\
\frac{{\vec c}}{\ell} \end{array}\right]
\left({\vec z}-\frac{\tau}{\omega\kappa}
\left(n+\frac{\beta}{2\pi}\right){\hat e}+
\frac{\alpha}{2\pi\omega\kappa}{\hat e},\Omega\right)
\end{equation}
where
$\Omega = 1/\ell\Omega '$.

We now express the functions $F_n$ in the above basis:
\begin{equation}
F_{n}\left({\vec z}\right) = \sum_{{\vec c}}D_{{\vec c},n}
\theta\left[\begin{array}{c}{\vec 0} \\
\frac{{\vec c}}{\ell} \end{array}\right]
\left({\vec z}-\frac{\tau}{\omega\kappa}
\left(n+\frac{\beta}{2\pi}\right){\hat e}+
\frac{\alpha}{2\pi\omega\kappa}{\hat e},\Omega\right)
\end{equation}
Imposing the quasi-periodicity conditions on the $F_n$
relate the coefficients:
\begin{equation}
D_{{\vec c},p-1} = D_{{\vec c}, 0}e^{-i2\pi\left(q\mid c \mid +
N_{\Phi}c_{\gamma}\right)/\ell} e^{\frac{i\pi\tau}{\omega\kappa^2}
\left(p-1\right)^{2}} e^{i\tau\beta N_{A}\left(p-1\right)/
{\omega\kappa^2}}
e^{-i\alpha N_{A}\left(p-1\right)/{\omega\kappa^2}}
\end{equation}
with the requirement:
\begin{equation}
e^{-i2\pi\left(q\mid c \mid +pN_{\Phi}c_{\gamma}\right)/\ell} = 1
\end{equation}
(where the notation $\mid c \mid = \sum_{\gamma = 1}^{N_A}
c_{\gamma}$ has been introduced) and
\begin{equation}
D_{{\vec c},n-1} = D_{{\vec c},
n}e^{-i2\pi N_{\Phi}c_{\gamma}/\ell}
e^{\frac{i\pi\tau}{\omega\kappa^2}}
e^{\frac{-i2\pi\tau N_{A}}{\omega\kappa^2}\left(n+\beta /2\pi\right)}
e^{i\alpha N_{A}/\omega\kappa^2}
\end{equation}
Further, diagonalizing the center of mass translation operators
$T\left({\vec L_2}\right)$ with eigenvalue $\exp i2\pi k_{2}/N$ gives
for $n < b$:
\begin{eqnarray}
D_{{\vec c},n+p-b} = D_{{\vec c},n} e^{-i2\pi k_{2}/N}
e^{-i2\pi \left(pw + qN\left(p-b\right)\right)\mid c \mid /\ell pN}
e^{\frac{i\pi\tau}{\omega\kappa^2} \left(p-b\right)^{2}}
e^{\frac{i2\pi\tau N_{A}}{\omega\kappa^2}\left(n+\beta/2\pi\right)
\left(p-b\right)}
e^{\frac{-i\alpha N_{A}}{\omega\kappa^2}\left(p-b\right) }
\end{eqnarray}
and for $n > b$:
\begin{equation}
D_{{\vec c},n-b} = D_{{\vec c},n} e^{-i2\pi k_{2}/N}
e^{-i2\pi \left(pw -qNb\right)\mid c \mid /\ell pN}
e^{\frac{i\pi\tau}{\omega\kappa^2}b^{2}}
e^{\frac{-i2\pi\tau N_{A}}{\omega\kappa^2}\left(n+\beta/2\pi\right)b}
e^{\frac{i\alpha N_{A}}{\omega\kappa^2}b}
\end{equation}
Diagonalizing $T\left({\vec L_1}\right)$ with eigenvalue
$\exp i2\pi k_{1}/N$ gives:
\begin{equation}
D_{{\vec c},n} = D_{{\vec c}+p^{2}r\omega{\hat e},n}e^{i2\pi k_{1}/N}
\end{equation}
Finally, requiring that the action of the translation operators be
independent of particle index i.e. $t_{\alpha} = t_{\beta}$
(identical particles) imposes the constraint:
\begin{equation}
\exp -i2\pi N_{\Phi}c_{\gamma}/\ell = \exp -i2\pi N_{\Phi}c_{\delta}/
\ell
\end{equation}
for any $\gamma , \delta$.

Using these relations one can express the eigenvalue
$\exp i2\pi k_{2}/N$ in terms of ${\vec c}$:
\begin{equation}
e^{i2\pi k_{2}/N} = e^{-i2\pi sN_{\Phi}c_{\gamma}/\ell}
e^{i2\pi \mid c\mid r/\ell}
\end{equation}
Notice that due to the constraints on ${\vec c}$ , the right hand
side is an Nth root of unity and is independent of the index $\gamma$.
Given ${\vec c}$ satisfying the above constraints the
coefficients can be solved for giving the most general form:
\begin{equation}
D_{{\vec c}, n} = d_{\vec c}
e^{\frac{i\pi}{\kappa^2}\left( n+ \frac{\beta}{2\pi}\right)^{2}{\hat
e}\cdot\Omega{\hat e}}  e^{-i2\pi \left(n/\kappa + \beta /2\pi\kappa
\right){\hat e}\cdot \left(\frac{{\vec c}}{\ell} +{\hat e}
\frac{\alpha}{2\pi\omega\kappa}\right)} e^{\frac{i2\pi n}{p\ell}
\left(q\mid c\mid + pN_{\Phi}c_{\gamma}\right)} e^{ i\frac{\beta\mid
c\mid}{\kappa \ell}}e^{i2\pi\frac{k_{1}\mid a \mid} {\ell N_{A}}}
\end{equation}
with the condition
\begin{equation}
d_{{\vec c} + r\omega p^{2}{\hat e}} = d_{\vec c}
\end{equation}
The coefficients $D_{{\vec c},n}$ satisfy the important property
of being symmetric under a permutation of the entries of the vector
${\vec c}$.

We now give a resolution of the constraints on $\vec {c}$ \footnote{
I thank Malin Ljungberg for her collaboration on this part of the
paper}.
The  vector $\vec {c}$ satisfies the constraints if and only if there
exist integers $m_{0},m_{\gamma}$ with $0 \leq m_{0} <p$ such that
the following equation is satisfied:
\begin{equation}
c_{\gamma} = pN_{\Phi}\left( m_{0} + pm_{\gamma}\right) +
\frac{q}{p\omega}\sum_{\delta=0}^{N_A}\left(c_{\gamma}-
c_{\delta}\right)
\end{equation}
Given
a solution ${\vec c}$ to the above equation, the eigenvalue $k_{2}$
can be expressed as:
\begin{equation}
k_{2}\bmod N = \frac{1}{p^{2}\omega}\sum_{\delta = 0}^{N_A}
\left(c_{\delta}-c_{\gamma}\right)
\end{equation}
and $k_1$ can be picked to be any integer in the range $0 \leq k_{1}
< N$ such that $k_{1}M = 0\bmod N$ where $M$ is defined to be the {\it
smallest} integer in the range $0 < M \leq N$ such that:
\begin{equation}
{\vec c} + Mrp^{2}\omega {\hat e} = {\cal P}{\vec c} + 0\bmod
{\ell Z^{N_A}}
\end{equation}
for some ${\cal P} \in S_{N_A}$.  That is, $M$ is the smallest integer
such that the left hand side is a permutation of ${\vec c}$ (with
the components of the vector defined $\bmod \ell$).

Let $A_{\gamma\delta}$ be the $N_{A}\times N_{A}$ matrix which acts
on a vector by exchanging its $\gamma$th and $\delta$th entry.
By a simple calculation one derives:
\begin{eqnarray}
&  &\theta\left[\begin{array}{c}{\vec 0} \\
\frac{{\vec c}}{\ell}  \end{array}\right]
\left(A_{\gamma\delta}{\vec z}-\frac{\tau}{\omega\kappa}
\left(n+\frac{\beta}{2\pi}\right){\hat e}+
\frac{\alpha}{2\pi\omega\kappa}{\hat e},\Omega\right) \nonumber \\
& = &\theta\left[\begin{array}{c}{\vec 0} \\
\frac{A_{\gamma\delta}{\vec c}}{\ell}  \end{array}\right]
\left({\vec z}-\frac{\tau}{\omega\kappa}
\left(n+\frac{\beta}{2\pi}\right){\hat e}+
\frac{\alpha}{2\pi\omega\kappa}{\hat e},\Omega\right)
\end{eqnarray}
Since any permutation can be written as a string of elementary
transpositions $A_{\gamma\delta}$; given ${\vec c}$
satisfying equation (42) and ordered ($0 \leq c_{1} \leq
\cdots \leq c_{N_A} < \ell$),
 we can symmetrize the functions $F_{n,k_{1},{\vec c}}$:
\begin{equation}
F_{n,k_{1},{\vec c}} = \sum_{{\cal P}\in S_{N_A}}\sum_{K=0}^{M-1}
D_{{\vec c}+Kp^{2}r\omega , n}
\theta\left[\begin{array}{c}{\vec 0} \\
\frac{{\cal P}{\vec c}}{\ell}+\frac{K}{N}{\hat e}
 \end{array}\right]
\left({\vec z}-\frac{\tau}{\omega\kappa}
\left(n+\frac{\beta}{2\pi}\right){\hat e}+
\frac{\alpha}{2\pi\omega\kappa}{\hat e},\Omega\right)
\end{equation}

We are in a position to state the complete set of basis functions for
$\psi$.  Again assume $\vec {c}$ satisfies equation (42) and that its
components are written in non-decreasing order, in addition assume
that $0 \leq c_{1} < rp^{2}\omega$ to prevent counting basis
elements more than once.  A basis for the wave functions is then
given by:

\begin{eqnarray}
& &\psi_{l,k_{1},{\vec c}} =
e^{-\frac{\pi\omega}{L_{1}L_{2}}\sum y_{\gamma}^{2}}
e^{Xv/L_1}\prod_{\gamma < \delta} \mid \theta_{1}\left( z_{\gamma}
-  z_{\delta} \mid \tau \right) \mid^{\frac{1}{\kappa}} \nonumber \\ &
& \sum_{K=0}^{M-1}\sum_{n=0}^{p-1}\sum_{{\cal P}\in S_{N_A}}
e^{i2\pi n\left(q\mid c \mid + pN_{\Phi}c_{\gamma}\right)/p\ell}
e^{i2\pi K\left(ns/\kappa+\beta s/2\pi\kappa +k_{2}/N\right)}
\nonumber \\ &  &
\theta\left[\begin{array}{c} -{\hat e}\left(n/\kappa \beta /2\pi\kappa
\right) \\  \left(K/N + {\alpha}/2\pi\omega\kappa\right){\hat e}
+{\cal P}{\vec c}/\ell
\end{array}\right]
\left({\vec z},\Omega\right)f_{ln}\left( v\right)
\end{eqnarray}
where $k_1$ satisfies the condition stated above ($k_{1}M =0\bmod N$)
and
\begin{equation}
f_{ln} = \sum_{j=-\infty}^{\infty}e^{i\alpha\left(l+jq+n/\kappa
\right)}\delta\left( \kappa v -2\pi\kappa l -2\pi\left(jp+n\right)
-\beta\right)
\end{equation}
An irrelevant overall factor has been discarded in the above
expression.

Next we turn to the special case of the exactly filled lowest
Landau level, $N_{\Phi} = 0$.  In this particular case it is
covenient to factorize the wave function as a product of
a center of mass and relative wave function. \footnote{Note that this
is not always possible.  It is only possible if $N_{\Phi}=0$ or
in the case $N_{\Phi}\neq 0$ when $s=1$, i.e. when $N_{\Phi}$ is
divisible by $N_A$.}  Write
\begin{equation}
F_{n}\left({\vec z}\right) = h_{n}\left(Z\right) g\left({\vec z}_{1}
\right)
\end{equation}
where $h_n$ is holomorphic in $Z = \left(X+iY\right)/L_{1}$,
$g$ is independent of the index $n$ and depends only on the
relative coordinates ${\vec z}_1$, $z_{1\gamma} = z_{1} - z_{\gamma}$.
The hamiltonian is periodic under translation along both non-trivial
cycles when $N_{\Phi}=0$ and hence the magnetic translation operators
become ordinary translation operators, and the wave functions
become exactly periodic.  We now prove that $g$ is a constant.
To prove this result, fix $\gamma\neq 1$ and consider the center
of mass preserving translations:$x_{\gamma} \rightarrow x_{\gamma}
+ L_{1}\left(N_{A}-1\right)$, $x_{\delta}
\rightarrow x_{\delta}-L_{1}$ ($\delta\neq\gamma$) and $y_{\gamma}
\rightarrow y_{\gamma} + L_{2}\left(N_{A}-1\right)$,
$y_{\delta}  \rightarrow y_{\delta}-L_{2}$($\delta\neq\gamma$).  Single
valuedness of $\psi$ under both translations gives:
\begin{equation}
g\left(z_{1\gamma}+N_{A},
\left\{z_{1\delta}\right\}_{\delta\neq\gamma}\right)
= g\left(z_{1\gamma},
\left\{z_{1\delta}\right\}_{\delta\neq\gamma}\right)
\end{equation}
\begin{equation}
g\left(z_{1\gamma}+N_{A}\tau,
\left\{z_{1\delta}\right\}_{\delta\neq\gamma}\right)
= g\left(z_{1\gamma},
\left\{z_{1\delta}\right\}_{\delta\neq\gamma}\right)
\end{equation}
That is, $g$ is a doubly periodic entire function in each of the
$z_{1\gamma}$.  By Liouville's first theorem $g$ is a
constant \cite{knopp}.  Periodicity of $\psi$ imposes constraints on
the $h_n$
\begin{equation}
h_{n}\left(Z+1\right) =h_{n}\left(Z\right)
\end{equation}
\begin{equation}
h_{n}\left(Z+\tau\right) =
e^{-i\pi\frac{\tau}{\kappa}} e^{i\frac{\tau}{\kappa}\left(2\pi n +
\beta \right)} e^{-i\frac{\alpha}{\kappa}}h_{n-1}\left(Z\right) \mbox{
$n>0$}
\end{equation}
\begin{equation}
h_{0}\left(Z+\tau\right) = e^{-i\pi\frac{\tau}{\kappa}}
e^{i\frac{\tau}{\kappa}\beta}
e^{-i\frac{\alpha}{\kappa}}e^{-i2\pi qZ}h_{p-1}\left(Z\right)
\end{equation}
which imply:
\begin{equation}
h_{n}\left( Z+p\tau\right) =e^{-i\pi\tau pq}
e^{iq\tau\left(2\pi n + \beta \right)}
e^{-iq\alpha}e^{-i2\pi qZ}h_{n}\left(Z\right)
\end{equation}
By an identical argument as the one presented for the $N_{\Phi}\neq 0$
case, the reader will easily convince herself that a basis for
$h_{n}$ is provided by:
\begin{eqnarray}
h_{n,c} & = & e^{i\pi\tau\kappa\left(\frac{n}{\kappa} +
\frac{\beta}{2\pi\kappa}\right)^2}e^{-i2\pi\left(\frac{n}{\kappa}+\frac{\beta}
{2\pi\kappa}\right)\left(\frac{\alpha}{2\pi}+\frac{c}{q}\right)}
e^{-i2\pi cn/p} \nonumber \\ & &\theta\left[\begin{array}{c} 0
\\ \frac{c}{q} +
\frac{\alpha}{2\pi}
\end{array}\right]
\left( Z-\tau\left(n+\beta /2\pi\right)+\alpha /2\pi, \tau\kappa\right)
\end{eqnarray}
where $c$ is an integer in the range $0\leq c < q$.
The $q$-fold degenerate set of wave functions can be written as:
\begin{eqnarray}
\psi_{l,c} & = & e^{-\pi\frac{\omega}{L_{1}L_{2}}\sum y_{\gamma}^2}
e^{Xv/L_1}\prod_{\gamma < \delta} \mid \theta_{1}\left( z_{\gamma}
-  z_{\delta} \mid \tau \right) \mid^{\frac{1}{\kappa}}
e^{iXv} \nonumber \\ & &\sum_{n=0}^{p-1} e^{i2\pi cn/p}
\theta\left[\begin{array}{c} -\left(n/\kappa
+\beta /2\pi\kappa \right)
\\ c/q + \alpha /2\pi \end{array}\right]
\left( Z, \tau\kappa\right) f_{ln}\left(v\right)
\end{eqnarray}
where $f_{ln}$ was defined above.  An irrelevant constant factor has
once again been discarded.

Finally, we perform a gauge transformation to bring the hamiltonian
to a form which describes anyons in an external magnetic field with
no long-range Aharonov-Bohm interactions.  Define the wave
function ${\tilde \psi}_{l}^{0}$ \cite{ans} as:
\begin{equation}
\psi_{l} =\prod_{\gamma < \delta}
\left(\frac{\theta_{1}^{\ast}\left( z_{\gamma} -
z_{\delta} \mid \tau \right)}{\theta_{1}\left( z_{\gamma} -
z_{\delta} \mid \tau \right)}
 \right)^{\frac{1}{2\kappa}} e^{iXv}
{\tilde \psi}_{l}^{0}
\end{equation}
This gauge transformation removes the long range interaction
between particles keeping only the interaction with the
external magnetic field.
${\tilde \psi}_{l}^{0}$ picks up the correct
phase when two particle coordinates are exchanged.  Let us
investigate how the wave function transforms under translations
by a lattice vector.  It will be convenient to define the following
$p$ set of wave functions, for $N_{\Phi}\neq 0$:
\begin{eqnarray}
& &{\tilde \psi}_{l,k_{1},{\vec c}}^{j} =
e^{-\frac{\pi\omega}{L_{1}L_{2}}\sum y_{\gamma}^{2}}
\prod_{\gamma < \delta} \left(\theta_{1}\left( z_{\gamma}
-  z_{\delta} \mid \tau \right)\right)^{\frac{1}{\kappa}} \nonumber \\
& &\sum_{K=0}^{M-1}\sum_{n=0}^{p-1}\sum_{{\cal P}\in S_{N_A}}
e^{-i2\pi\frac{nj}{\kappa}}
e^{i2\pi n\left(q\mid c \mid + pN_{\Phi}c_{\gamma}\right)/p\ell}
e^{i2\pi K\left(ns/\kappa+\beta s/2\pi\kappa +k_{2}/N\right)}
\nonumber \\ & &
\theta\left[\begin{array}{c} -{\hat e}\left(n/\kappa \beta /2\pi\kappa
\right) \\  \left(K/N + {\alpha}/2\pi\omega\kappa\right){\hat e}
+{\cal P}{\vec c}/\ell
\end{array}\right]
\left({\vec z},\Omega\right)f_{ln}\left( v\right)
\end{eqnarray}
and for $N_{\Phi}=0$
\begin{eqnarray}
\psi_{l,c} & = & e^{-\pi\frac{\omega}{L_{1}L_{2}}\sum y_{\gamma}^2}
\prod_{\gamma < \delta} \left( \theta_{1}\left( z_{\gamma}
-  z_{\delta} \mid \tau \right) \right)^{\frac{1}{\kappa}}
e^{iXv} \nonumber \\
& &\sum_{n=0}^{p-1} e^{i2\pi cn/p}e^{-i2\pi\frac{jn}{\kappa}}
\theta\left[\begin{array}{c} -\left(n/\kappa
+\beta /2\pi\kappa \right)
\\ c/q + \alpha /2\pi
\end{array}\right]
\left( Z, \tau\kappa\right) f_{ln}\left(v\right)
\end{eqnarray}
Now under a translation $x_{\alpha}\rightarrow x_{\alpha} + L_1$,
while holding the other coordinates fixed, the wave function
changes its component index as follows:
\begin{equation}
{\tilde \psi}_{l}^{j}\left(x_{\alpha}+L_{1}\right)
= e^{-i\frac{\beta}{\kappa} + i\frac{\pi}{\kappa}\left(N_{A}-1\right)}
{\tilde \psi}_{l}^{j+1}\left(x_{\alpha}\right)
\end{equation}
A translation $y_{\alpha}\rightarrow y_{\alpha} + L_2$ does not
affect the component index of the wave function but it picks up
a component dependent phase:
\begin{equation}
{\tilde \psi}_{l}^{j}\left(y_{\alpha}+L_{1}\right)
= e^{i2\pi\omega x_{\alpha}/L_{1}}
e^{i2\pi\frac{j}{\kappa} -
i\frac{\pi}{\kappa}\left(N_{A}-1\right)} {\tilde
\psi}_{l}^{j+1}\left(y_{\alpha}\right)
\end{equation}
Where the degeneracy indices have been suppressed.
There are three important points to note here: 1) the wave function
picks up the correct gauge transformation under translation since
now only the external magnetic field is present, the C-S contribution
has been "gauged" away; 2) the wave function has $p$ components;
3) the components survive even when $\omega$ is itself an integer
(that is, the components are not an artifact of the non-commutativity
of the appropriate magnetic translation operators).
Finally, it is interesting to observe that when $N_{\Phi}=0$ the ground
state is non-degenerate only when $q=1$.  So statistics of the type
$\pi /p$ are special in that they have non-degenerate ground states
when the filling factor is exactly one.  Moreover, the $N_{\Phi}=0$
states are of the Laughlin-Jastrow form investigated by Haldane and Rezayi
in the fermion case \cite{hr}.

To summarize, we have found the complete set of ground
state wave functions for $N_{A}/\omega\kappa \leq 1$.  The wave
functions have $p$ components consistent with both arguments
based on the braid group and the statistics transmuting
Chern-Simons field \cite{ein}.  Though we have provided a complete basis of
wave functions we have been unable to give a closed expression for
the total degeneracy for the case $N_{\Phi}\neq 0$.  This is due
to the rather complicated constraints satisfied by the vector $\vec
{c}$.  A more transparent resolution of the
constraint may be possible allowing one to give a closed expression
for the degeneracy.

I thank T. H. Hansson, A. Karlhede, M. Ljungberg
and E. Westerberg for illuminating discussions.

\end{document}